\documentclass[12pt]{revtex4}
\usepackage{epsfig}
\usepackage{graphicx}
\usepackage{dcolumn}
\usepackage{bm}
\usepackage[hang,bf,small]{caption}
\usepackage{amsmath}
\DeclareGraphicsExtensions{.eps.gz,.eps,.ps,.ps.gz}
\parskip=\itemsep               

\begin{document}
\setcounter{footnote}{1}
\title{Saturation 2005 ( mini-review)}

\author{Eugene Levin}\email[]{leving@post.tau.ac.il} \email[]{levin@mail.desy.de}

\affiliation{
{\it HEP Department}\\
{\it School of Physics and Astronomy}\\
{\it Raymond and Beverly Sackler Faculty of Exact Science}\\
{\it Tel Aviv University, Tel Aviv, 69978, ISRAEL}\\
}

\centerline{}

\begin{abstract}

This talk is a brief review of ups and downs of high density QCD during the past year.
\end{abstract}

\maketitle


\section{ Saturation at HERA and RHIC}
Before discussing the high density QCD news we would like to summarize what we have  learned about 
saturation at  HERA and RHIC.

  HERA:

\begin{itemize}
\item  \quad The power - like growth of $x G(x,Q^2)$ at low $x$ ($x G(x,Q^2) \propto x^{- \lambda}$ 
with $\lambda \approx 0.3$;
\item \quad The geometrical scaling behaviour for $x \leq 10^{-2}$;
\item \quad Fit of  all HERA data for $Q^2 = 0 \div 500 \,GeV^2$ with 
 $\chi^2/d.o.f. \,\,\leq\,1$\, based on non-linear equation \cite{Levin:GLLM,Levin:IIM};
\end{itemize}

  RHIC:

\begin{itemize}
  \item \quad  Saturation approach for  $d N/d y$ versus $y$, energy and number of 
participants  predicted  and led to a reasonable description of  the experimental data \cite{Levin:KLN};
\item \quad  Prediction for suppression of the hadron production in dA 
collision
and confirmation  in the experimental data \cite{LEVIN:KLM,Levin:KKT}.
\end{itemize}

The only consistent explanation all these observations is to assume that at HERA we have started to
approach a new phase of QCD,  with large gluon  density  but still with small coupling 
constant. 
 The regime of high parton density at HERA is reached due to  the
QCD emission of gluons that was incorporated in the QCD evolution equations.
The independent check of the effects of high gluon density at HERA  was performed 
by RHIC
experiment in heavy ion-ion collisions. In this reaction the energies are much lower than at HERA,  
but
the large values of the parton densities were  achieved due to the large number of nucleons in a 
nucleus. Based on these experimental observations we can anticipate that the  LHC will be a machine 
for 
discovery  a new phase of QCD: colour glass condensate with saturated   gluon density.

\section{ Predictions for the LHC range of energies }

Our main challenge is to provide reliable estimates for the influence of high density QCD 
(saturation) effects in  the LHC range of energies. The first such estimates have been discussed 
\cite{Levin:GLMN,Levin:ESQU},  and the results  for the ratio of the unintegrated 
structure functions  $D = \phi^{NL}/\phi^{L}$ are   plotted in Fig.1 where 
\begin{equation} \label{Levin:inc}
 \frac{d \sigma}{d y d^2 p_t} \,\, \propto \,\frac{\alpha_S}{p^2_t}\,\int\,d^2
k_t\,\phi(k^2_t)\,\,\phi(
(\vec{p} - \vec{k})^2_t)
\end{equation}
and $\phi^{NL} ( \phi^{L})$ is solution of the non-linear (linear ) equation.  

\begin{figure}
\includegraphics[width=0.98\textwidth]{{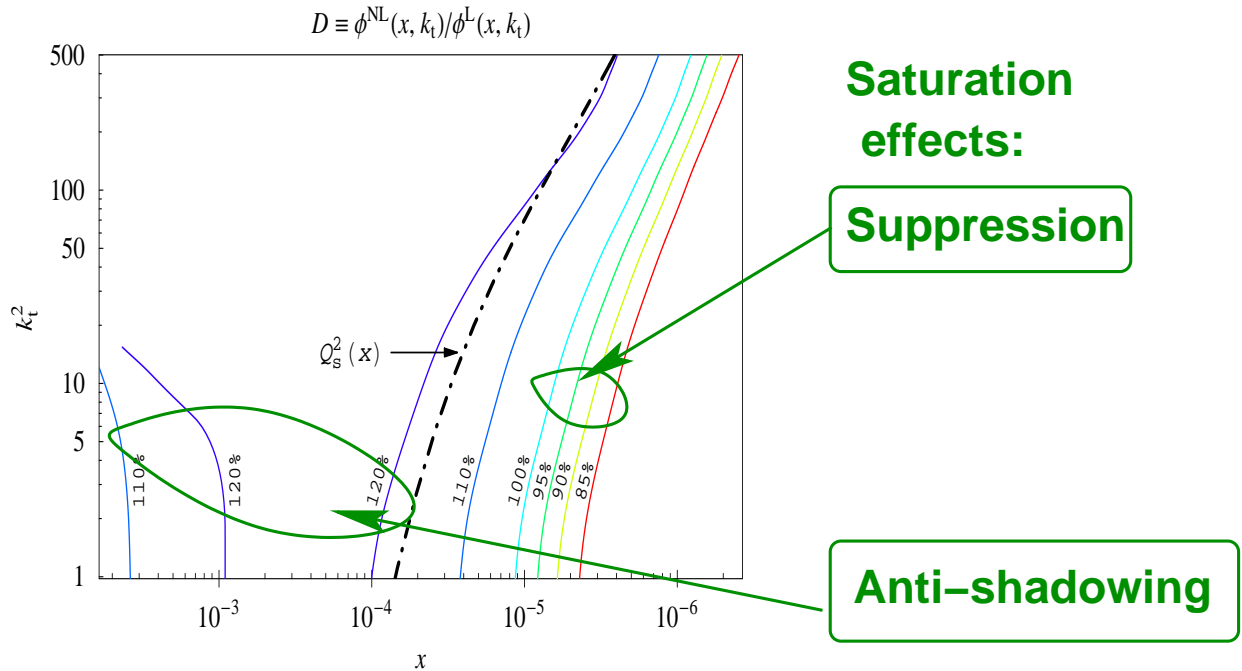}}
\caption{The ratio $\phi^{NL}/\phi^{L}$ which shows the influence of non-linear correction on the 
predictions for inclusive gluon jet production at LHC energies}
\end{figure}
It should be stressed that non-linear evolution predicts not only suppression in the saturation 
region, but also the anti-shadowing effect which results in an increase of the value of $\phi$ for 
$Q^2 > 
Q^2_s(x)$,  where $Q_s$ is the saturation scale. One can see that the suppression and  increase 
could 
be rather large leading to an inclusive cross section twice as large or twice as small, as  the 
predictions based on  routine linear evolution. 

\section{Theoretical development}
\subsection{ B- JIMWLK approach $\mathbf{\longleftrightarrow}$  BFKL Pomeron  Calculus}
The good news is that it turns out that Balitsky-JIMWLK approach \cite{Levin:JIMWLK}
can be reduced to BFKL Pomeron calculus \cite{Levin:KL}, and JIMWLK effective Lagrangian give us 
possibility to calculate all multi-Pomeron vertices. For the first time, we can do such calculations 
using operator formalism without spending years to obtain result just summing Feyman diagrams.    
Since the colour dipoles are the `wee' partons of the BFKL equation the  Balitsky-JIMWLK formalism 
can be 
discussed in terms of the dipole approach. 

The bad news is that we have not  achieved any progress in Pomeron calculus.

\subsection{ Probabilistic interpretation}
Our last hope is the probabilistic approach to Pomeron interaction. The best way to express our
optimism is to cite  Grassberger and  Sundermeyer \cite{Levin:GS} who proposed this interpretation:
{\it `` Reggeon field theory is  equivalent to a chemical process
where a radical can undergo diffusion, absorption, recombination, and autocatalytic production.
Physically, these "radicals" are wee partons ({\em colour dipoles})}".

It turns out that B-WLKJIM approach can be written as a typical death-birth process (Markov's 
chain)\cite{Levin:BIW,Levin:LL}
\begin{equation} \label{Levin:MC}
\frac{\partial P_n}{\partial Y}\, =\,- \sum_i \Gamma(1 \to 2) 
\bigotimes\,\left( P_n (...x_i,y_i ...;Y) \, -\,  P_{n-1} (...x_i,y_i ...;Y) \right)
\end{equation}
where $P_n$ - probability to find $n$-dipoles at rapidity $Y$, $\Gamma(1 \to 2)$ describe the decay 
of one dipole into two dipoles and $\bigotimes$ denotes all needed integration. This equation can be 
a basis for the Monte Carlo code which will be able to solve high density QCD equations, and 
which will 
lead  to theoretical treatment of the multiparticle production.

\subsection{Hunt for Pomeron loops}
The process of two Pomeron  merging into one Pomeron is naturally included in Pomeron calculus with 
the same vertex as the process of Pomeron splitting. However, we need correctly normalize this 
process if we wish to use the probabilistic interpretation. Such normalization was suggested in Ref. 
\cite{Levin:IT} and this vertex $\Gamma(2 \to 1)$ has been calculated 
\cite{Levin:IT,Levin:MSW,Levin:LL}.  Using this vertex, we can generalize Eq.(2) which takes the 
form
\begin{equation} \label{Levin:GPN}
 \frac{\partial P_n}{\partial Y}\, =\,\mbox{Eq.(2)} \,- \sum_i \Gamma(2 \to 1)
\bigotimes\,\left( P_n (...x_i,y_i ...;Y) \, -\,  \sum_k P_{n+1} (...x_i,y_i ...x_k,y_k;Y) \right)
\end{equation}
\subsection{Solution}
Attempts  to solve Eqs.(3) have been made  in Refs.\cite{Levin:BO,Levin:L,Levin:RS}. The result is 
surprisingly unexpected, namely,
\begin{itemize}
\item \quad  Asymptotic solution leads to a {\em gray} disc ({\em not black!!!});
\item \quad   Using the large parameters of our theory  ($\Gamma(1\to 2)/\Gamma(2\to 1) \approx 
N^2_c/\alpha^2_S$ and $ \Gamma(1\to 2)/\Gamma(2\to 3) \approx N^2_c$)  the
 semiclassical approach can be developed  for searching for both the asymptotic solution and the
corrections to it, at high energy;
\item \quad  The corrections to the asymptotic solution decrease at large values of $Y$, and can be
found from  the  Liouville-type linear equation;
\item \quad  The important role in searching for high energy asymptotic behaviour of the amplitude 
plays the role of  $t$-channel   unitarity constraint, which specifies the value of the typical 
amplitude for 
dipole-dipole interaction.
\end{itemize}
\subsection{Topics which I have no room to discuss}
This brief review is my personal view on news in low $x$ (high density) QCD. Unfortunately, I had no 
room even to express my point of view. It is pity since I think that a more microscopic approach, 
related to the new effective Lagrangian,  and to a search for a  Bogolubov transformation between 
dipole 
and quarks (antiquark) and gluon degrees of freedom \cite{Levin:KL,Levin:HIMST,Levin:MMSW}, looks 
very interesting. It is very attractive approach and I hope that my references provide the  reader 
with 
names of active players in this field. However, I must admit that the  theory becomes dangerously 
complicated and reminds me more and more my nightmare  that Lipatov \cite{Levin:LI} is correct with 
his effective 
action,  which is not easier to solve than the full QCD Lagrangian. 

{\bf  Acknowledgments:} I am very grateful to E. Gotsman for everyday discussions on the subject of 
this talk. 

\end{document}